\documentclass[11pt]{article}
\usepackage[T1]{fontenc}
\usepackage[utf8]{inputenc}
\usepackage{amsmath,amssymb,amsthm}
\usepackage{graphicx}
\usepackage{geometry}
\usepackage[colorlinks,linkcolor=blue,citecolor=blue,urlcolor=blue]{hyperref}
\hypersetup{pdftitle={voigtinference: Exact likelihood calculus and
conditional attribution for the Voigt profile},
pdfauthor={Peter Reinhard Hansen and Chen Tong},
pdfkeywords={Voigt profile, Faddeeva function, maximum likelihood,
Fisher information, deconvolution, line-shape fitting}}
\usepackage{booktabs}
\usepackage[section]{placeins}
\usepackage{xcolor}

\newtheorem{theorem}{Theorem}
\theoremstyle{remark}
\newtheorem{remark}{Remark}

\newcommand{\E}{\mathbb{E}}
\newcommand{\V}{\mathbb{V}}
\newcommand{\reOp}{\operatorname{Re}}
\newcommand{\imOp}{\operatorname{Im}}

\begin{document}

\title{\textsf{voigtinference}: Exact likelihood calculus and conditional attribution for the Voigt profile\thanks{%
Corresponding author:
Peter Reinhard Hansen (hansen@unc.edu). We thank Sebastian Ament for helpful
discussions and for independently benchmarking alternative numerical
methods for evaluating the Voigt profile. Chen Tong acknowledges financial support from the Youth Fund of the
National Natural Science Foundation of China (72301227) and the Fujian
Provincial Natural Science Foundation of China (2025J08008).}}
\author{Peter Reinhard Hansen\\
{\small Department of Economics, University of North Carolina at Chapel Hill}
\and Chen Tong\\
{\small School of Economics, Xiamen University}}
\date{21 August 2026}
\maketitle

\begin{abstract}
Fast, accurate algorithms for the Faddeeva function $w(z)$, and hence for the
Voigt profile $K(x,a)$, have existed for four decades, and analytic first
derivatives are also available in some implementations. What the
established libraries reviewed here have not provided is the full
likelihood calculus of the normalized Voigt
\emph{distribution}: applications still commonly resort to pseudo-Voigt
approximations, finite-difference derivatives, or numerical convolution for
inference on the profile's parameters. Because $w'(z)=-2z\,w(z)+2i/\sqrt\pi$,
every derivative of the Voigt log-likelihood is an algebraic function of $K$ and
the dispersion part $L(x,a)=\imOp\,w(z)$, obtained from the same single complex
evaluation that delivers the profile. This yields the score and Hessian in
closed form, and the expected Fisher information by one-dimensional quadrature
of an analytic integrand; for fixed interior widths $\sigma,\gamma>0$, maximum
likelihood estimation of the center and both
widths is consistent and asymptotically normal at rate $\sqrt n$, despite the
distribution having neither a finite mean nor a finite variance, so
conventional likelihood-based standard errors apply. The
conditional mean of the Gaussian component given an observation is
$(y-\mu)-\gamma L/K$: a redescending function that attributes moderate deviations
to the Gaussian (Doppler/resolution) component and extreme ones to the Lorentzian
tail. The package \textsf{voigtinference} (Python, NumPy/SciPy, with a
cross-validated Julia companion) supplies the unified toolkit: score, full
parameter Hessian, expected Fisher information, Newton-based unbinned
maximum likelihood with boundary diagnostics, conditional component
moments, and evaluation validated against high-precision references
across extreme width ratios. It applies directly to unbinned non-relativistic, constant-width
Breit--Wigner\,$\otimes$\,Gaussian resonance fits and supplies analytic
Jacobians for line-shape refinement.
\end{abstract}

\section*{SOFTWARE SUMMARY}
\begin{small}
\noindent
\emph{Program Title:} voigtinference 1.1.1\\
\emph{Developer's repository link:} \url{https://github.com/reinhardhansen/voigtinference}\\
\emph{Licensing provisions:} MIT\\
\emph{Programming language:} Python ($\geq$ 3.10; depends only on NumPy and
SciPy, whose \texttt{scipy.special.wofz} supplies the Faddeeva primitive). A
companion Julia implementation (\textsf{VoigtInference.jl}, Julia $\geq$ 1.10,
SpecialFunctions.jl) is included and agrees within $10^{-12}$ on all
evaluation-level quantities over the distributed validation grid (every
difference exactly zero in the authors' recorded environment).\\
\emph{Supplementary material:} cross-language benchmark and
$10^{-12}$-tolerance cross-check (\texttt{bench/}), high-precision validation
of the numerical
branches (\texttt{examples/certify.jl}), and scripts reproducing every
numerical table and figure in this paper.\\[4pt]
\emph{Nature of problem:} Likelihood-based inference for the Voigt profile,
the convolution of a Gaussian (Doppler/resolution) and a Lorentzian
(lifetime/pressure) component. Existing software evaluates the profile, and
in some cases its first derivatives, but not the statistical calculus of the
normalized distribution. Required are: unbinned maximum likelihood estimation
of the center and both widths from event-level data; asymptotic standard
errors from the expected or observed information; diagnostics for the
boundary cases in which one component is undetected; and the conditional
moments of the latent Gaussian component given a noisy observation
(deconvolution/attribution). The distribution has no finite positive integer
moments, so conventional estimation based on positive integer moments is
unavailable, and naive derivative formulas lose
all significant digits deep in the Lorentzian tail and at extreme width
ratios.\\[4pt]
\emph{Solution method:} The identity $w'(z)=-2z\,w(z)+2i/\sqrt{\pi}$ makes all
derivatives of the Voigt log-likelihood rational functions of $K(x,a)$ and
$L(x,a)$, obtained from one Faddeeva evaluation per data point. The package
assembles the analytic score and Hessian, evaluates the expected Fisher
information by one-dimensional quadrature of the analytic score, and maximizes
the likelihood by safeguarded Newton iteration in $(\mu,\log\sigma,\log\gamma)$
with projected-gradient convergence tests, boundary-submodel diagnostics, and
positive-definiteness-checked standard errors; conditional moments follow from
Tweedie's formula in closed form. Cancellation-prone regimes are dispatched to
extensively validated Cauchy-limit expansions gated on
$r=\sigma^2/((y-\mu)^{2}+\gamma^{2})$.\\[4pt]
\emph{Additional comments including restrictions and unusual features:} The
non-relativistic, constant-width Breit--Wigner\,$\otimes$\,Gaussian
(``Voigtian'') shape used in unbinned event-level fits is the special case
with center equal to the resonance mass; analytic gradients can be passed to
MINUIT-style optimizers, and a fused density-plus-score routine supplies
least-squares Jacobians. Restrictions: the likelihood theory assumes iid data
and interior widths $\sigma,\gamma>0$; Wald standard errors are suppressed
(reported as NaN) when the maximizer is on or effectively on a width
boundary, where the interior asymptotics do not apply; the optimizer is a
local Newton method with optional deterministic multistart, not a verified
global maximizer. Double-precision CPU implementation (no GPU support); numerical
branches validated on a grid spanning $\gamma/\sigma\in[10^{-8},10^{8}]$. The Raman example
reads a small spectrum distributed with a third-party CRAN package (not
redistributed here); all other examples are self-contained with fixed seeds.
\end{small}

\bigskip
\noindent\textbf{Keywords:} Voigt profile; Faddeeva function; maximum likelihood;
Fisher information; deconvolution; line-shape fitting

\section{Introduction}\label{sec:intro}

The Voigt profile, the convolution of a Gaussian and a Lorentzian, describes
spectral lines under combined Doppler and pressure/lifetime broadening,
diffraction peaks in powder analysis, quasar absorption lines, and, as a
Breit--Wigner resonance smeared by Gaussian detector resolution, mass peaks in
particle physics. Its \emph{evaluation} is a solved problem: the profile is the
real part of the Faddeeva function, for which fast, accurate algorithms have been
available since Huml\'{\i}\v{c}ek \cite{Humlicek:1982}, with later refinements in
\cite{ZaghloulAli:2011}, related high-accuracy work for the wider family of
stable densities \cite{AmentONeil:2018}, and implementations in standard
libraries (\texttt{scipy.special.voigt\_profile} \cite{Virtanen:2020},
\texttt{RooVoigtian} \cite{VerkerkeKirkby:2003}).

\emph{Inference} for the profile's parameters is less well served. Some
building blocks exist: Astropy's \texttt{Voigt1D} model exposes analytic first
derivatives of the line shape for least-squares fitting through its
\texttt{fit\_deriv} mechanism \cite{Astropy:2022}, ROOT's \texttt{RooVoigtian}
supports unbinned likelihood fits (with gradients supplied numerically or,
recently, by automatic code generation) \cite{BrunRademakers:1997,VerkerkeKirkby:2003},
and the analytic \emph{argument} derivatives of the Voigt function itself are
classical (Section~\ref{sec:density}). But among the widely used packages
reviewed here, none supplies the statistical calculus of the normalized
Voigt \emph{distribution}: the likelihood score and
full parameter Hessian, the expected Fisher information, maximum likelihood
with boundary diagnostics, and the conditional moments of the latent
components. Applied work therefore still commonly uses pseudo-Voigt
approximations \cite{ThompsonCoxHastings:1987,IdaAndoToraya:2000}, least
squares with finite-difference Jacobians \cite{Newville:2014}, or
simulation-based methods \cite{CannasPiras:2025}, and unbinned likelihood
fits commonly rely on numerical gradients inside the optimizer.
Table~\ref{tab:software} summarizes the landscape.

\begin{table}[htb]
\centering
\caption{Voigt-profile capabilities of widely used software. ``analytic
$\partial_\theta$'' means analytic first derivatives with respect to the
line-shape parameters; conditional moments refer to the latent
Gaussian/Lorentzian attribution of Section~\ref{sec:cond}. Entries reflect
the cited documentation as accessed in August 2026, conservatively; the
claim is about the packages reviewed here, not all software.}
\label{tab:software}
\begin{footnotesize}
\begin{tabular}{lcccccc}
\toprule
 & profile & analytic $\partial_\theta$ & Hessian & Fisher info & unbinned MLE & cond.\ moments \\
\midrule
SciPy \cite{Virtanen:2020} & \checkmark & & & & & \\
Astropy \texttt{Voigt1D} \cite{Astropy:2022} & \checkmark & \checkmark & & & & \\
ROOT \texttt{RooVoigtian} \cite{BrunRademakers:1997,VerkerkeKirkby:2003} & \checkmark & & & & \checkmark & \\
lmfit \texttt{VoigtModel} \cite{Newville:2014} & \checkmark & & & & & \\
VoigtFit \cite{Krogager:2018} & \checkmark & & & & & \\
CRAN \texttt{voigt} \cite{CannasPiras:2025} & \checkmark & & & & & \\
\textsf{voigtinference} (this paper) & \checkmark & \checkmark & \checkmark & \checkmark & \checkmark & \checkmark \\
\bottomrule
\end{tabular}
\end{footnotesize}
\end{table}

This paper provides the missing toolkit. The derivative identity for the
Faddeeva function is classical; what it implies statistically is that the
score, Hessian, and conditional moments of the Gaussian
component are all algebraic in the real and imaginary parts of the \emph{same}
complex evaluation that delivers the profile, and that the expected Fisher
information reduces to one-dimensional quadrature of these expressions.
The division of labor with the companion paper \cite{HansenTong:2026} is as
follows. The analytic identities and the asymptotic MLE theorem below are
proved there; this paper contributes their formulation in
line-shape/Faddeeva conventions, the reference implementations
(\textsf{voigtinference} in Python, the archived program, and
\textsf{VoigtInference.jl} in Julia, its cross-validated companion), the
numerical-stability design and its high-precision validation, the
cross-language benchmarks, and the finite-sample and spectroscopic
validation, all of which are new here.

\section{Notation and density}\label{sec:density}

Let $Z\sim\mathcal N(0,\sigma^2)$ and $X\sim\operatorname{Cauchy}(0,\gamma)$
(Lorentzian with HWHM $\gamma$) be independent, and let
\begin{equation}
Y=\mu+Z+X,\qquad \theta=(\mu,\sigma,\gamma)'.
\end{equation}
The density of $Y$ is the Voigt profile. With the Faddeeva function
$w(z)=e^{-z^2}\operatorname{erfc}(-iz)$ and
\begin{equation}
z=x+ia,\qquad x=\frac{y-\mu}{\sigma\sqrt2},\qquad a=\frac{\gamma}{\sigma\sqrt2},
\end{equation}
write $K=K(x,a)=\reOp\,w(z)$ and $L=L(x,a)=\imOp\,w(z)$ for the absorption and
dispersion parts. Then \cite{Kendall_D:1938}
\begin{equation}\label{eq:density}
f_Y(y;\theta)=\frac{K(x,a)}{\sigma\sqrt{2\pi}}.
\end{equation}
Two classical facts drive everything below: $K(x,a)>0$ everywhere, and
\begin{equation}\label{eq:wprime}
w'(z)=-2z\,w(z)+\frac{2i}{\sqrt\pi},
\end{equation}
so $w$ is closed under differentiation. Using this closure to obtain
\emph{argument} derivatives of the Voigt function is itself classical:
Heinzel \cite{Heinzel:1978} derived recurrence relations for the partial
derivatives with respect to $(x,a)$ to arbitrary order, and Wells
\cite{Wells:1999} computes $\partial K/\partial x$ and $\partial K/\partial a$
simultaneously with the function for use as retrieval Jacobians; Webb, Carswell
and Lee \cite{WebbCarswellLee:2021} recently developed analytic Voigt
derivatives specifically to replace finite differences in high-resolution
absorption-line fitting. What appears not to have been developed systematically
is the corresponding likelihood calculus for the normalized Voigt
\emph{distribution}: every
finite-order derivative of $\log f_Y$ with respect to the \emph{parameters}
$(\mu,\sigma,\gamma)$ (and $y$) is a rational function of $x$, $a$, $K$, and
$L$, with denominators that are powers of $K>0$ and $\sigma>0$: one Faddeeva
evaluation per data point yields the density and, algebraically, the
likelihood calculus (score, Hessian, and conditional moments; the expected
Fisher information then requires only one-dimensional quadrature). In particular, analytic Jacobians and
Hessians for \emph{least-squares} line-shape fitting come from the same pair
$(K,L)$, replacing finite differences in Levenberg--Marquardt refinement.

\section{Likelihood inference}\label{sec:mle}

Write $\tilde y=y-\mu$. The score of $\log f_Y(y;\theta)$ is
\cite{HansenTong:2026}
\begin{equation}\label{eq:score}
s_{\mu}=\frac{\tilde y-\gamma\frac{L}{K}}{\sigma^2},\qquad
s_{\sigma}=\frac{(\tilde{y}^{2}-\gamma^{2}-\sigma^{2})K-2\gamma\tilde{y}L
  +\sqrt{\tfrac{2}{\pi}}\,\sigma\gamma}{\sigma^{3}K},\qquad
s_{\gamma}=\frac{\gamma K+\tilde{y}L-\sqrt{\tfrac{2}{\pi}}\,\sigma}{\sigma^{2}K},
\end{equation}
and the Hessian assembles recursively from the scores and $L/K$, with no further
special-function calls:
\begin{equation}\label{eq:hessian}
\begin{aligned}
H_{\mu\mu} &= \tfrac{1}{\sigma}s_{\sigma}-s_{\mu}^{2}, &
H_{\mu\sigma} &= -\tfrac{1}{\sigma}\big(s_{\mu}+\gamma H_{\mu\gamma}-\tilde{y}H_{\mu\mu}\big),\\
H_{\gamma\gamma} &= -\tfrac{1}{\sigma}s_{\sigma}-s_{\gamma}^{2}, &
H_{\gamma\sigma} &= -\tfrac{1}{\sigma}\big(s_{\gamma}+\gamma H_{\gamma\gamma}-\tilde{y}H_{\mu\gamma}\big),\\
H_{\mu\gamma} &= \tfrac{1}{\sigma^{2}}\big(\tilde{y}s_{\gamma}+\gamma s_{\mu}-\tfrac{L}{K}\big)-s_{\mu}s_{\gamma}, &
H_{\sigma\sigma} &= -\tfrac{1}{\sigma}\big(s_{\sigma}+\gamma H_{\gamma\sigma}-\tilde{y}H_{\mu\sigma}\big).
\end{aligned}
\end{equation}
The Faddeeva form eliminates the exponential instability of differentiating
the convolution integral directly, and $K>0$ everywhere means no denominator
can vanish. Positivity does not by itself prevent floating-point
cancellation: the numerators of these algebraic formulas lose digits deep in
the Lorentzian tail and at extreme width ratios, which is handled by the
validated asymptotic branches described in Section~\ref{sec:software}. A
fused likelihood-score-Hessian evaluation requires one complex evaluation
per observation, the cost of evaluating the profile itself.

The Voigt distribution has no finite positive integer moments (the Lorentzian
tail dominates, so neither the mean nor the variance exists), but the score and
information are perfectly regular, and likelihood asymptotics are conventional:

\begin{theorem}[{Hansen and Tong \cite{HansenTong:2026}}]\label{thm:mle}
Let $Y_1,\ldots,Y_n$ be iid $\,\mathcal V(\mu_0,\sigma_0,\gamma_0)$ and let
$\hat\theta_n$ maximize the log-likelihood over a compact
$\Theta\subset\mathbb R\times(0,\infty)^2$ with $\theta_0$ interior. Then
$\hat\theta_n\xrightarrow{p}\theta_0$, the information-matrix equality
$\mathcal I_{\theta_0}=\E[s\,s']=-\E[H]$ holds with $\mathcal I_{\theta_0}$
nonsingular, and
$\sqrt n\,(\hat\theta_n-\theta_0)\xrightarrow{d}
\mathcal N(0,\mathcal I_{\theta_0}^{-1})$, with
$\mathcal I_{\mu_0\sigma_0}=\mathcal I_{\mu_0\gamma_0}=0$ by symmetry.
\end{theorem}

Conventional likelihood-based standard errors, $t$-statistics, and confidence
intervals therefore apply: observed-information standard errors follow directly
from the analytic Hessian, and expected-information standard errors from
one-dimensional quadrature of the analytic score (Section~\ref{sec:software}).
Theorem~\ref{thm:mle} is an interior result; when the maximizer effectively
lies on the Gaussian boundary $\gamma\to0$, as happens in small samples with
weak Lorentzian components, the Wald/Fisher approximation is not expected to
apply (Section~\ref{sec:numerics}). The center estimate is asymptotically
independent of the width
estimates, and the ratio of the asymptotic standard deviations of
$(\hat\sigma,\hat\gamma)$ depends only on $\gamma/\sigma$; the Lorentzian width is
the more precisely estimated over the empirically relevant range, being identified
by the tails while $\sigma$ is identified in the core \cite{HansenTong:2026}.

\begin{remark}[Unbinned resonance fits]
A non-relativistic, constant-width Breit--Wigner resonance of mass $M$ and
full width $\Gamma$ under Gaussian resolution $\sigma$ has event-level density
$\mathcal V(M,\sigma,\Gamma/2)$.
Eqs.~\eqref{eq:score}--\eqref{eq:hessian} supply analytic gradients and Hessians
for MINUIT-style optimizers \cite{JamesRoos:1975}, and Theorem~\ref{thm:mle} the associated asymptotic
standard errors for $(M,\sigma,\Gamma)$. The derivatives require no
automatic-differentiation machinery and reuse the Faddeeva value already
computed for the density.
\end{remark}

\section{Deconvolution: conditional moments of the Gaussian component}\label{sec:cond}

Given an observation $Y=y$, how much of the deviation is Gaussian
(Doppler/resolution) and how much Lorentzian? Since the model is an additive
Gaussian convolution, Tweedie's formula \cite{Robbins:1956,Efron:2011},
$\E[Z|Y=y]=-\sigma^2\partial_y\log f_Y(y)$, applies and gives closed forms
\cite{HansenTong:2026}:
\begin{equation}\label{eq:cond}
\E[Z|Y=y]=\tilde y-\gamma\frac{L}{K},
\qquad
\V(Z|Y=y)=\sqrt{\tfrac{2}{\pi}}\,\frac{\sigma\gamma}{K}
-\gamma^{2}\Big(1+\frac{L^{2}}{K^{2}}\Big),
\end{equation}
with $\E[X|Y=y]=\gamma L/K$, well defined despite $\E|X|=\infty$. The
attribution is governed entirely by the dispersion-to-absorption ratio $L/K$.
Near the center the conditional mean is approximately linear in $\tilde y$, as in
Gaussian signal extraction; far in the tails, where the Lorentzian dominates,
$L/K\to\tilde y/\gamma$ and $\E[Z|Y=y]\to0$: the update is
\emph{redescending}, so deviations too large to be Gaussian are attributed to the
Lorentzian component automatically. Its extrema occur where
$\V(Z|Y=y)=\sigma^2$. Higher conditional cumulants follow from further
derivatives of $\log f_Y$. In \cite{HansenTong:2026} this exact static
conditional mean is the update function of a Masreliez-type
\cite{Masreliez:1975} \emph{approximate} filter for time series with Voigt
measurement noise: because the Gaussian prediction law there is an
approximation, the recursive estimator is quasi-maximum-likelihood, distinct
from the exact iid likelihood inference treated here; that development is not
repeated.

\section{The software: \textsf{voigtinference} (Python) and
\textsf{VoigtInference.jl} (Julia)}\label{sec:software}

The methods are provided in two implementations with parallel APIs for the
principal quantities: a Python package, \textsf{voigtinference} (NumPy
\cite{Harris:2020} and SciPy \cite{Virtanen:2020} only), which is the
archived program, and a Julia \cite{Bezanson:2017} package,
\textsf{VoigtInference.jl} (SpecialFunctions.jl only), in which the software
was developed. On the distributed validation grid, spanning both algorithmic
branches, three width ratios $\gamma/\sigma\in\{10^{-4},0.58,10^{4}\}$, and
$|\tilde y|/\sqrt{\sigma^2+\gamma^2}$ up to $10^{8}$, the distributed
cross-check requires the two implementations to agree within $10^{-12}$ on
every evaluation-level quantity (density, log-density, score, Hessian, and
both conditional moments) and fails otherwise; in the authors' recorded
environment every such difference is exactly zero. Achieving that requires
matching not just the algebra but the association order of the
floating-point expressions, which is why bitwise agreement is reported as
an observed property of one environment rather than promised across
compilers and special-function libraries. Both implement
Eqs.~\eqref{eq:density}--\eqref{eq:cond} with one complex evaluation per
observation.

Two implementation details matter. First, the exact expressions lose digits
to floating-point cancellation whenever the Gaussian component is a small
perturbation of the Cauchy component: deep in the tail (e.g.\ $\tilde yL$
and $\sqrt{2/\pi}\,\sigma$ in $s_\gamma$ agree to many digits while their
difference is the answer), but also at \emph{any} $\tilde y$ when
$\gamma\gg\sigma$, where the digits lost grow like
$4\log_{10}(\gamma/\sigma)$ even at the line center. Both regimes are
covered by a single criterion: the implementation switches to the
Cauchy-limit expansions, obtained from
$f=c+\tfrac{\sigma^2}{2}c''+O(\sigma^4c^{(4)})$ with $c$ the Lorentzian
density, wherever the expansion parameter
$r=\sigma^2/(\tilde y^{2}+\gamma^{2})$, which controls their relative error,
is small. Second, the Cauchy-limit branches are not ad hoc pointwise
approximations: every score and Hessian entry is the exact derivative of
\emph{one} truncated expansion of the log density,
$\log f=\log c+\tfrac{\sigma^2}{2}A+\tfrac{\sigma^4}{8}(B-A^2)
+\sigma^6(\tfrac{C}{48}-\tfrac{AB}{16}+\tfrac{A^3}{24})$ with
$A=c''/c$, $B=c^{(4)}/c$, $C=c^{(6)}/c$. This construction matters for
inference, not only for evaluation: a pointwise-accurate branch whose
entries are not derivatives of a common objective can violate the
likelihood identities after integration, because the leading term cancels
under the expectation while a truncation bias does not. A first-order
branch, for instance, yields $\E[s_\sigma]\gamma^4/\sigma^3\to\tfrac12$
instead of zero and makes $-\E[H]$ indefinite at large $\gamma/\sigma$.
The consistent construction restores $\E[s]=0$ and $\E[ss']=-\E[H]$ to
the retained order $O(r^3)$, and the distributed test suites
verify both identities by quadrature under the exact model, together with
positive definiteness of $-\E[H]$. Because the branch is then more
accurate than the exact formulas well before their cancellation becomes
visible, the thresholds sit early: $r<10^{-4}$ for the score and
conditional moments and $r<5\times10^{-4}$ for the Hessian, the values
minimizing the worst-case error among the candidates of the validation
scan. In the band $10^{-4}\leq r<5\times10^{-4}$ the public score
consequently uses the exact formulas while the public Hessian uses the
expansion; each branch is internally derivative-consistent, and the
dispatched outputs agree with the exact derivatives within the validated
worst-case bounds below. The dispatched
implementation is validated against high-precision references on a grid
spanning $\gamma/\sigma\in[10^{-8},10^{8}]$, including line centers, branch
crossovers, and the first actually-dispatched floating-point neighbor on
each side of every threshold, with normwise worst-case errors (largest component
error over the block's largest component) of $1.4\times10^{-10}$ for the
score, $6.3\times10^{-7}$ for the Hessian, $1.6\times10^{-12}$ for the
conditional mean, and $6.2\times10^{-12}$ for the conditional variance,
which is nonnegative on the entire grid; the validation driver
(\texttt{examples/certify.jl}), whose high-precision reference is itself
checked by an internal two-method overlap comparison, is distributed with the
packages and exits nonzero on any violation. Main
entry points (identical in both implementations):

\begin{center}\small
\begin{tabular}{ll}
\toprule
Function & Purpose \\
\midrule
\texttt{voigt\_pdf}, \texttt{voigt\_logpdf} & density / log-density \eqref{eq:density} \\
\texttt{voigt\_score}, \texttt{voigt\_hessian} & Eqs.~\eqref{eq:score}, \eqref{eq:hessian} \\
\texttt{voigt\_pdf\_score} & density + score from one pass (LS Jacobians) \\
\texttt{voigt\_fisher} & Fisher information $\mathcal I_\theta$ \\
\texttt{voigt\_mle} & MLE by safeguarded Newton; both s.e.\ sets \\
\texttt{boundary\_lr} & closed-family boundary LR statistics \\
\texttt{voigt\_condmean}, \texttt{voigt\_condvar} & conditional moments \eqref{eq:cond} \\
\bottomrule
\end{tabular}
\end{center}

Optimization is Newton's method with the analytic score and Hessian in
$(\mu,\log\sigma,\log\gamma)$ coordinates, safeguarded by a ridge and a
projected backtracking line search (the Armijo condition is tested on the
executed, clamped step), from moment-free starting values (median for $\mu$;
tail and interquartile quantiles for the width split, since the Voigt
distribution has no finite positive integer moments and moment-based
initialization is unavailable),
with an optional deterministic multistart. Convergence is declared on the
projected gradient, and the fit reports its termination reason. Convergence is
typically reached in five to seven Newton iterations. The result carries the
boundary diagnostics used in Section~\ref{sec:numerics}: separate flags for
each width, set when the corresponding lower log-width clamp is active,
when that width falls below $10^{-6}$ times the other, or when the fitted
log likelihood exceeds the corresponding nested-submodel log likelihood by
no more than $10^{-7}$ (the log-$\gamma$ gradient decays linearly as
$\gamma\to0$, whereas the log-$\sigma$ gradient is quadratic in $\sigma$
through $\tau=\sigma^{2}$, so an effective boundary need not touch the
literal clamp); closed-form Gaussian and Newton-based Cauchy boundary-submodel fits
with their log-likelihoods, so the boundary likelihood-ratio diagnostic of
Section~\ref{sec:mc} is always available; and standard errors from both the
expected information (Gauss--Legendre quadrature after a tangent substitution
that maps the Lorentzian tails to a bounded integrand) and the observed
information, each checked for positive definiteness and suppressed (reported
as NaN, with flags) at a boundary, where the interior asymptotics of
Theorem~\ref{thm:mle} do not apply. The test suite validates every formula against central
finite differences, the Tweedie identities, the information-matrix equality,
and parameter recovery on simulated data, in both languages. The optimizer
evaluates the log-likelihood, score, and Hessian in a single fused pass (one
Faddeeva evaluation per observation per iteration), and the Gauss--Legendre
nodes for the Fisher quadrature are computed once and cached. Automatic differentiation is unnecessary here; in the NumPy/SciPy and Julia
SpecialFunctions stacks used by these packages, it is also not off-the-shelf,
because generic forward-mode AD does not compose through the complex
special-function evaluation without custom derivative rules (frameworks with
such rules exist elsewhere, e.g.\ ROOT's automatic gradient generation). The
analytic expressions \eqref{eq:score}--\eqref{eq:hessian} sidestep the issue
entirely and reuse the Faddeeva value already computed for the density.

A minimal session, from \texttt{examples/demo.py}, fits a simulated sample
of $n=5{,}000$ draws from $\mathcal V(0.5,\,1.0,\,0.3)$ (the summary line
is wrapped to fit the column):

\begin{small}
\begin{verbatim}
>>> from voigtinference import rand_voigt, voigt_mle
>>> y = rand_voigt(5_000, 0.5, 1.0, 0.3, rng=2026)
>>> r = voigt_mle(y)
>>> print(r.summary())
Voigt MLE   n-iterations=7  converged=True
            termination=gradient_converged  loglik=-9818.059090
             estimate    std. error     obs. s.e.
     mu      0.515466      0.018786      0.018790
  sigma      1.014695      0.023145      0.023171
  gamma      0.293235      0.017507      0.017510
boundary submodels: loglik(Gaussian) = -20019.447246,
                    loglik(Cauchy)   = -10136.486406
\end{verbatim}
\end{small}

The fit costs eight Faddeeva passes over the sample, and the accounting is
inclusive: one pass per accepted Newton iteration (the line search's
evaluation at the accepted point is reused; rejected trials would add
passes but none occurred here), plus one pass for the observed information
after convergence. The boundary-submodel log-likelihoods confirm an
interior Voigt fit (the likelihood-ratio statistics against the Gaussian
and Cauchy submodels are far beyond the boundary critical value of
Section~\ref{sec:mc}). Every estimate is within one standard error of the
truth
($t$-statistics $+0.82$, $+0.63$, $-0.39$), and the expected- and
observed-information standard errors agree to three digits. The fitted
conditional moments illustrate the redescending attribution of
Section~\ref{sec:cond}: $\E[Z|Y=y]$ rises from $0$ at $y=\hat\mu$ to a
maximum of $1.40$ (that is, $1.38\,\hat\sigma$) near $y-\hat\mu\approx2.4$,
then redescends through $0.49$ at $y-\hat\mu=5$, $0.21$ at $10$, and $0.02$
at $100$; $\V(Z|Y=y)$ equals $\hat\sigma^2$ at the peak, rises above it in
the ambiguous region ($1.50$ at $y-\hat\mu=3$), and returns to
$\hat\sigma^2$ ($1.0296$) as the observation is fully attributed to the
Lorentzian component ($\V=1.0298$ at $y-\hat\mu=100$).

\section{Numerical validation}\label{sec:numerics}

We validate the implementation in six ways, one per subsection below, all
reproducible from scripts distributed with the packages (each table and
figure caption names its script). Every script uses fixed seeds and prints
its expected outputs, and the benchmark cross-check and the validation
driver exit nonzero on any violation. The fixed seeds make the
simulations and every validation decision reproducible in the stated
software environments; the benchmark timings themselves remain machine-
and load-dependent.

\subsection{Monte Carlo: finite-sample estimation and coverage}\label{sec:mc}

Table~\ref{tab:mc} reports bias, RMSE, and coverage of 95\%
Fisher-information intervals for sample sizes
$n\in\{10^2,10^3,10^4,10^5\}$ and width ratios
$\lambda=\gamma_0/\sigma_0\in\{0.01,0.1,1\}$. The grid spans the physically
relevant regimes: $n\sim10^2$ corresponds to discovery-level event counts in
unbinned resonance fits, $n\sim10^3$--$10^4$ to routine resonance fits and
photon-counting line spectroscopy, and $n\sim10^5$ to precision calibration
samples; $\lambda=0.01$ is the canonical narrow-resonance problem of
extracting a Lorentzian width far below the Gaussian resolution, while
$\lambda\approx1$ corresponds to strongly mixed profiles. When the Cauchy
component is weakly identified (small $\lambda$ and small $n$), the maximizer
can run toward the Gaussian limit $\gamma\to0$. Operationally, the optimizer
works in $(\mu,\log\sigma,\log\gamma)$ with the log-widths clamped at
$10^{\pm8}$ times a quantile-based initial width scale, and ``boundary''
means the fit is numerically Gaussian by any of three criteria: the lower
clamp on $\log\gamma$ is active; $\hat\gamma<10^{-6}\hat\sigma$ (the
log-width gradient vanishes with the width -- like $\gamma$ here and like
$\sigma^{2}$ on the other boundary -- so an effective boundary need not
touch the clamp); or the fitted likelihood exceeds the Gaussian submodel's
by less than $10^{-7}$ log-likelihood units, a likelihood-equivalent fit
being that submodel whatever the fitted width says. The mirrored criteria
diagnose an effectively Cauchy fit. The implementation reports this rather
than failing, and it
suppresses the Wald standard errors there, since the interior asymptotics of
Theorem~\ref{thm:mle} do not apply on the boundary. Coverage in
Table~\ref{tab:mc} is accordingly unconditional: a sample with suppressed
standard errors counts as non-covering. This is the operationally relevant
number, and it makes the cost of weak identification visible: at
$\lambda=0.01$, $n=10^2$, the Cauchy component goes undetected in $77\%$ of
samples, reported interior Wald intervals exist in the remaining $23\%$,
and unconditional
coverage is $0.21$--$0.23$ for all three parameters, recovering to
$0.79$--$0.81$ at $n=10^3$ (boundary share $18\%$) and to within $0.015$ of
nominal from $n=10^4$ on. Conditional on a reported interval, coverage at
the corner design is $0.945$ ($\mu$), $0.906$ ($\sigma$), $0.997$ ($\gamma$):
conditioning flatters the intervals precisely where they are least
available. The same attenuation appears in milder form at $\lambda=0.1$,
$n=10^2$ (boundary share $15\%$, unconditional coverage $0.80$--$0.83$),
and on the opposite boundary at $\lambda=1$, $n=10^2$, where $9\%$ of fits
are effectively Cauchy ($\hat\sigma$ undetected) and unconditional coverage
is $0.85$--$0.86$.

The boundary flag alone, however, understates how often the data are
consistent with a pure submodel, because on (or near) a boundary the width
estimate frequently sits at a small interior value rather than on the
boundary itself. For $\sigma$ this has a clean local description: in
$\tau=\sigma^2\geq0$ the boundary score is
$\partial_\tau\log f|_{\tau=0}=\tfrac12\,c''/c$, whose variance under the
Cauchy null, $\E_c[(c''/c)^2]/4=1/(4\gamma^4)$, is finite and remains
positive after projecting out the nuisance parameters (the efficient
boundary information is
$\mathcal I_{\tau\tau\cdot(\mu,\gamma)}=1/(8\gamma^4)>0$), so standard
one-sided theory applies: $\hat\tau=\max(0,O_p(n^{-1/2}))$, the
closed-family MLE has $\hat\sigma=0$ in about half of all samples and
$\hat\sigma=O_p(n^{-1/4})$ otherwise, and the null law of
$2(\ell_{\mathcal V}-\ell_{\mathrm{Cauchy}})$ approaches the
$\tfrac12\chi^2_0+\tfrac12\chi^2_1$ mixture with $5\%$ point $2.706$
\cite{SelfLiang:1987}. At the Gaussian boundary $\gamma=0$ no such
description is available: the $\gamma$-score at $\gamma=0$ behaves as
$\sqrt{2/\pi}\,(\sigma/\tilde y^{2})\,e^{\tilde y^{2}/2\sigma^{2}}$ in the
tail, the Fisher information is infinite, and we
claim no rate or limit law there. On both sides the reliable diagnostic is
the likelihood ratio against the fitted boundary submodel, computed over
the \emph{closed} family: the closed Voigt family contains both submodels,
so the full-model likelihood is the maximum of the returned multistart
interior candidate and the two fitted submodels (the optimizer is a local
method, so this is the computed closed-family comparison, not a
certificate of the global interior supremum), and
\[
\mathrm{LR}_{\mathrm{sub}}
=\max\bigl\{0,\;2\bigl(\max(\ell_{\mathcal V},\ell_{\mathrm G},
\ell_{\mathrm C})-\ell_{\mathrm{sub}}\bigr)\bigr\}
\]
is nonnegative by construction (\texttt{boundary\_lr} in both packages).
The raw difference against the interior candidate alone is not a
likelihood ratio: near a boundary that candidate legitimately sits a clamp
residual below the exact submodel likelihood. Both LR
statistics are location-scale pivotal under their nulls (the Voigt family
and each submodel are location-scale families and the implemented fitting
map is equivariant), so their exact finite-sample null distributions
depend only on $n$; the package calibrates the cutoffs once per
$n$ by simulation and validates their achieved size on independent
replications (\texttt{examples/boundary\_lr.jl}).
The calibration ($9999$ replications per $n$; tail-probability standard
error about $0.002$) confirms both diagnoses: the Cauchy-side cutoffs are
$3.01$, $2.77$, $2.67$, $2.70$ for $n=10^2,\dots,10^5$, with the naive
$2.706$ rule having estimated finite-sample size $4.9$--$5.8\%$, while
the Gaussian-side
cutoffs are $0.107$, $0.048$, $0.004$, $0$, so the naive rule is
there not liberal but extremely conservative (estimated size at most
approximately $0.8\%$).
Under the Gaussian null the simulations place $93.6\%$ to $95.6\%$ of the mass
exactly at zero (the closed-family maximum is the Gaussian fit itself), so
the calibrated rule is approximately of level $5\%$ while the atom stays
below $95\%$, and conservative once the atom absorbs the quantile: at
$n=10^5$ the atom is $95.6\%$ and the cutoff is exactly zero. Its small
positive critical values are less sharply determined than their
Cauchy-side counterparts; the operative check is the achieved size on
independent validation replications, which is $0.047$, $0.052$, $0.049$,
$0.043$ on the Gaussian side and $0.046$--$0.050$ on the Cauchy side
(Monte Carlo standard error $0.002$).
The last two columns of Table~\ref{tab:mc} report retention at the
calibrated cutoffs: the Gaussian submodel is retained in $79\%$ of samples
at the corner design, decaying to zero by $n=10^4$, and the Cauchy submodel
is retained in $59\%$ of samples at $\lambda=1$, $n=10^2$; with a hundred
observations, a Gaussian core of width equal to the Lorentzian width is
hard to detect.

Biases are negligible from $n=10^3$ on and reach a few percent of the true
widths at $n=10^2$ (e.g.\ $-0.062$ for $\sigma$ at $\lambda=1$); RMSEs
scale as $1/\sqrt n$; with $5000$ replications the Monte Carlo noise in
each coverage entry is about $\pm0.006$ (two Monte Carlo standard errors), and in every design with zero
boundary share all coverage entries lie within $0.015$ of the nominal
level. In every design with zero boundary share, every fit converged on
the projected
gradient; at the weak-identification corners a fraction of fits ($33\%$ at
$\lambda=0.01$, $n=10^2$, falling to zero by $n=10^4$) instead terminate
in the numerically flat region at the $\gamma$ boundary, where no
representable likelihood improvement exists, and are reported as stalled
rather than converged; their boundary flags, submodel likelihoods, and
submodel standard errors carry the inference. The expected information was
positive definite in every one of the $60{,}000$ fits. Consistent with the Fisher-information analysis in
\cite{HansenTong:2026}, the Lorentzian width is estimated more precisely
than the Gaussian width once identified (e.g.\ at $\lambda=1$, $n=10^4$:
RMSE $0.027$ versus $0.039$).

\begin{table}[htb]
\centering
\caption{Finite-sample behavior of the estimates produced by the
exact-likelihood implementation: bias, RMSE, and
overall coverage of 95\% Fisher-information intervals, where a sample
whose intervals are suppressed at a boundary is counted as non-covering;
5000 replications per design with seven-start optimization, $\mu_0=0$,
$\sigma_0=1$, $\gamma_0=\lambda\sigma_0$.
$\mathrm{bdry}_{\mathrm G}$/$\mathrm{bdry}_{\mathrm C}$ are the shares of
effectively Gaussian ($\hat\gamma$ undetected) and effectively Cauchy
($\hat\sigma$ undetected) fits, in percent, and
$\mathrm{LR}_{\mathrm G}$/$\mathrm{LR}_{\mathrm C}$ the shares in which
the closed-family boundary likelihood-ratio test (\texttt{boundary\_lr})
at the simulation-calibrated nominal-$5\%$ cutoff (achieved sizes
reported in the text) retains each submodel;
definitions and calibration in
Section~\ref{sec:mc}. Produced by \texttt{examples/montecarlo.jl}.}
\label{tab:mc}
\begin{footnotesize}
\setlength{\tabcolsep}{3pt}
\begin{tabular}{rr rrr rrr rrr rrrr}
\toprule
 & & \multicolumn{3}{c}{$\mu$} & \multicolumn{3}{c}{$\sigma$} & \multicolumn{3}{c}{$\gamma$} & & & & \\
\cmidrule(lr){3-5}\cmidrule(lr){6-8}\cmidrule(lr){9-11}
$\lambda$ & $n$ & bias & RMSE & cov. & bias & RMSE & cov. & bias & RMSE & cov. & $\mathrm{bdry}_{\mathrm G}$ & $\mathrm{bdry}_{\mathrm C}$ & $\mathrm{LR}_{\mathrm G}$ & $\mathrm{LR}_{\mathrm C}$ \\
\midrule
0.01 & $10^2$ & $-$0.0006 & 0.1019 & 0.215 & $-$0.0108 & 0.0786 & 0.206 & 0.0011 & 0.0234 & 0.227 & 77.2 & 0.0 & 78.6 & 0.0 \\
0.01 & $10^3$ & 0.0005 & 0.0315 & 0.786 & $-$0.0005 & 0.0246 & 0.787 & 0.0000 & 0.0076 & 0.810 & 17.5 & 0.0 & 18.1 & 0.0 \\
0.01 & $10^4$ & 0.0001 & 0.0102 & 0.946 & 0.0000 & 0.0077 & 0.949 & 0.0000 & 0.0023 & 0.938 & 0.0 & 0.0 & 0.0 & 0.0 \\
0.01 & $10^5$ & 0.0000 & 0.0032 & 0.949 & 0.0000 & 0.0024 & 0.953 & 0.0000 & 0.0007 & 0.951 & 0.0 & 0.0 & 0.0 & 0.0 \\
\addlinespace
0.10 & $10^2$ & 0.0014 & 0.1140 & 0.798 & $-$0.0048 & 0.1168 & 0.801 & $-$0.0034 & 0.0713 & 0.830 & 15.4 & 0.0 & 16.2 & 0.2 \\
0.10 & $10^3$ & $-$0.0005 & 0.0357 & 0.947 & $-$0.0009 & 0.0349 & 0.947 & 0.0002 & 0.0220 & 0.942 & 0.0 & 0.0 & 0.0 & 0.0 \\
0.10 & $10^4$ & 0.0001 & 0.0111 & 0.945 & $-$0.0003 & 0.0110 & 0.948 & 0.0001 & 0.0071 & 0.950 & 0.0 & 0.0 & 0.0 & 0.0 \\
0.10 & $10^5$ & 0.0000 & 0.0035 & 0.951 & 0.0001 & 0.0034 & 0.949 & 0.0000 & 0.0022 & 0.945 & 0.0 & 0.0 & 0.0 & 0.0 \\
\addlinespace
1.00 & $10^2$ & 0.0072 & 0.2135 & 0.863 & $-$0.0621 & 0.4567 & 0.861 & $-$0.0209 & 0.2560 & 0.852 & 0.0 & 8.7 & 0.0 & 58.7 \\
1.00 & $10^3$ & $-$0.0001 & 0.0676 & 0.950 & $-$0.0038 & 0.1279 & 0.948 & $-$0.0033 & 0.0859 & 0.941 & 0.0 & 0.0 & 0.0 & 0.2 \\
1.00 & $10^4$ & 0.0005 & 0.0211 & 0.950 & $-$0.0005 & 0.0394 & 0.949 & 0.0000 & 0.0266 & 0.946 & 0.0 & 0.0 & 0.0 & 0.0 \\
1.00 & $10^5$ & $-$0.0001 & 0.0066 & 0.947 & 0.0000 & 0.0123 & 0.953 & $-$0.0001 & 0.0083 & 0.951 & 0.0 & 0.0 & 0.0 & 0.0 \\
\bottomrule
\end{tabular}
\end{footnotesize}
\end{table}

\subsection{Computational cost}\label{sec:cost}

Table~\ref{tab:timing} reports the per-observation cost of exact inference,
measured independently in both implementations. The qualitative cost
comparisons are similar across languages: the analytic score costs
$1.01$--$1.06\times$ the log-density evaluation alone (the practical content
of the claim that one Faddeeva evaluation delivers the density and all
derivatives), whereas central finite differences cost $\approx6\times$ for
the score (six likelihood evaluations in total) and $22$--$24\times$ for
the Hessian (24 evaluations); automatic differentiation through these stacks
requires custom rules (Section~\ref{sec:software}). The fused
likelihood-score-Hessian pass is nearly identical across the two
implementations ($97.0$ versus $97.2$ ns), while stand-alone routines
exhibit language-specific overheads (log-density $78.4$ versus $97.0$ ns;
stand-alone Hessian $97.6$ versus $171.5$ ns). Re-running after
minor-version updates of Python, Julia, and their special-function libraries
moved every entry by under $6\%$. Full maximum likelihood on $n=10^5$
observations, including both sets of standard errors, the boundary-submodel
fits, and all diagnostics of Section~\ref{sec:software}, takes $0.079$\,s in
Python and $0.146$\,s in Julia (6 Newton iterations each; the parameter
estimates agree across languages to
$\leq5\times10^{-14}$, at the level of the convergence tolerance; the Python
driver is faster end to end because it additionally reuses the line search's
Faddeeva evaluation at the accepted step, so an accepted iteration costs
exactly one pass over the sample).

\begin{table}[htb]
\centering
\caption{Per-observation cost of exact likelihood quantities for the Voigt
profile ($n=10^5$, single thread, Apple M1 Max; Python 3.14.7 with NumPy
2.5.2/SciPy 1.18.0, Julia 1.12.7 with SpecialFunctions 2.9.0; ratios to the
respective log-density in parentheses). Produced by
\texttt{bench/run\_bench.sh}.}
\label{tab:timing}
\begin{small}
\begin{tabular}{lrr}
\toprule
ns/observation & Python & Julia \\
\midrule
Faddeeva primitive                       & 71.9          & 85.1 \\
log-density                              & 78.4 (1.00)   & 97.0 (1.00) \\
analytic score, Eq.~\eqref{eq:score}     & 83.1 (1.06)   & 98.3 (1.01) \\
analytic Hessian, Eq.~\eqref{eq:hessian} & 97.6 (1.25)   & 171.5 (1.77) \\
fused log-lik.\ + score + Hessian        & 97.0 (1.24)   & 97.2 (1.00) \\
finite-difference score (6 evals)        & 470.5 (6.00)  & 558.6 (5.76) \\
finite-difference Hessian (24 evals)     & 1893.2 (24.2) & 2127.8 (21.9) \\
\bottomrule
\end{tabular}
\end{small}
\end{table}

\subsection{Tail and extreme-ratio accuracy}\label{sec:tailacc}

Table~\ref{tab:tail} illustrates the branch behavior of
Section~\ref{sec:software} for two representative designs: the dispatched
implementation reaches machine precision beyond
$|\tilde y|\sim10^{3}\sqrt{\sigma^2+\gamma^2}$ and is never worse than
$2.2\times10^{-12}$ in either design, while the naive formulas reach
normwise errors of order $10^{14}$--$10^{16}$. The
validation grid is broader: \texttt{examples/certify.jl} measures
normwise errors (largest component error over the block's largest true
component) against two-method high-precision references over
$\gamma/\sigma\in[10^{-8},10^{8}]$, including line centers, branch
crossovers, and exact-threshold and next-float points, giving the worst
cases quoted in Section~\ref{sec:software}. The naive double-precision
formulas, by contrast, lose all significant digits in the Hessian by
$|\tilde y|\sim10^{4}$--$10^{5}\sqrt{\sigma^2+\gamma^2}$ and reach normwise errors of
order $10^{14}$--$10^{16}$ at $10^{8}$ (Table~\ref{tab:tail}); at large
$\gamma/\sigma$ the loss sets in at the line center itself, which is why the
branches are gated on $r$ rather than on $|\tilde y|$. Constructing the references
themselves requires care for the same reason the switches exist: a relative
truncation error of order $\operatorname{erfc}(T)$ in the reference
quadrature is amplified by $\sim\tilde y^4$ through the cancellation, and
the required working precision grows with $|\tilde y|$: at
$|\tilde y|\sim10^{8}$ a 256-bit Hessian reference retains only $\sim$14
correct digits, so the Hessian columns are validated against a 512-bit
reference (the score columns need only 256 bits).

\begin{table}[htb]
\centering
\caption{Validation of the far-tail implementation: normwise error (defined
in the text) of the score and of the Hessian in double precision, for the
packaged implementation (Cauchy-limit branches gated on
$r=\sigma^2/(\tilde y^{2}+\gamma^{2})$ with $r_s=10^{-4}$,
$r_h=5\times10^{-4}$) and for the naive exact formulas without the
switches. References: 256-bit for the score, 512-bit for the Hessian, via
the integral representation of $\operatorname{erfcx}$; the generator
(\texttt{bench/tailtable.py}) includes a 256-versus-512-bit self-check.
The packaged columns are stable across platforms; the naive columns are
platform-sensitive (they measure the rounding of a cancelling expression,
which varies with the special-function library), so their exact values are
illustrative. Full validation grid over $\gamma/\sigma\in[10^{-8},10^{8}]$:
\texttt{examples/certify.jl}.}
\label{tab:tail}
\begin{small}
\begin{tabular}{r rr rr}
\toprule
 & \multicolumn{2}{c}{score} & \multicolumn{2}{c}{Hessian} \\
\cmidrule(lr){2-3}\cmidrule(lr){4-5}
$|\tilde y|/\sqrt{\sigma^2+\gamma^2}$ & package & naive & package & naive \\
\midrule
\multicolumn{5}{l}{$(\sigma,\gamma)=(1,1)$}\\
$10^1$ & $1.2\times10^{-14}$ & $1.2\times10^{-14}$ & $1.8\times10^{-12}$ & $1.8\times10^{-12}$ \\
$10^2$ & $3.2\times10^{-14}$ & $6.0\times10^{-12}$ & $2.2\times10^{-13}$ & $3.2\times10^{-7}$ \\
$10^3$ & $6.1\times10^{-17}$ & $5.3\times10^{-10}$ & $1.3\times10^{-16}$ & $3.0\times10^{-3}$ \\
$10^4$ & $2.8\times10^{-17}$ & $3.0\times10^{-9}$  & $3.9\times10^{-17}$ & $7.6\times10^{-1}$ \\
$10^5$ & $8.2\times10^{-18}$ & $7.3\times10^{-7}$  & $8.3\times10^{-18}$ & $2.9\times10^{4}$ \\
$10^6$ & $8.9\times10^{-17}$ & $2.8\times10^{-4}$  & $1.3\times10^{-16}$ & $9.5\times10^{8}$ \\
$10^7$ & $8.0\times10^{-18}$ & $1.0\times10^{0}$   & $8.0\times10^{-18}$ & $2.0\times10^{14}$ \\
$10^8$ & $1.0\times10^{-16}$ & $1.0\times10^{0}$   & $1.2\times10^{-16}$ & $4.0\times10^{16}$ \\
\addlinespace
\multicolumn{5}{l}{$(\sigma,\gamma)=(1,0.01)$}\\
$10^1$ & $1.1\times10^{-12}$ & $1.1\times10^{-12}$ & $2.2\times10^{-12}$ & $2.2\times10^{-12}$ \\
$10^2$ & $5.1\times10^{-15}$ & $9.7\times10^{-13}$ & $3.5\times10^{-16}$ & $9.7\times10^{-11}$ \\
$10^3$ & $2.2\times10^{-17}$ & $2.0\times10^{-10}$ & $1.5\times10^{-16}$ & $2.0\times10^{-6}$ \\
$10^4$ & $7.0\times10^{-17}$ & $1.9\times10^{-8}$  & $3.8\times10^{-17}$ & $1.9\times10^{-2}$ \\
$10^5$ & $1.2\times10^{-16}$ & $4.7\times10^{-7}$  & $2.3\times10^{-16}$ & $4.7\times10^{1}$ \\
$10^6$ & $6.3\times10^{-17}$ & $2.2\times10^{-6}$  & $2.4\times10^{-17}$ & $2.2\times10^{4}$ \\
$10^7$ & $2.3\times10^{-17}$ & $3.0\times10^{-2}$  & $4.0\times10^{-17}$ & $3.0\times10^{10}$ \\
$10^8$ & $2.1\times10^{-17}$ & $1.0\times10^{0}$   & $4.2\times10^{-17}$ & $1.0\times10^{14}$ \\
\bottomrule
\end{tabular}
\end{small}
\end{table}

\subsection{Pseudo-Voigt width distortion}\label{sec:pv}

The left panel of Figure~\ref{fig:figs} quantifies the parameter distortion
incurred when pseudo-Voigt parameters are interpreted as convolution widths.
The primary comparison uses the standard Thompson--Cox--Hastings
parameterization \cite{ThompsonCoxHastings:1987}: a Lorentzian and a
Gaussian of \emph{common} FWHM $\Gamma$ mixed with weight $\eta$, fitted by
maximum likelihood with free $(\mu,\Gamma,\eta)$ to $n=2\times10^5$ Voigt
draws per design, after which the fitted $(\Gamma,\eta)$ are mapped to
implied component FWHMs by inverting the Thompson--Cox--Hastings
width/mixing relations, the standard practice in powder diffraction. The
implied Gaussian width is overstated by $2.5$--$5.1\%$ across
$\lambda\in[0.25,4]$, and the implied Lorentzian width understated by
$10.2\%$ at $\lambda=0.25$, shrinking to $0.6\%$ at $\lambda=4$, whereas
the exact MLE recovers both widths to within $1.3\%$ throughout (sampling
noise at this $n$). A \emph{free}
Gaussian--Lorentzian mixture, $(1-\eta)\,\phi(y;\mu,\sigma_p)
+\eta\,c(y;\mu,\gamma_p)$ with independent widths, nests the
common-FWHM shape and fits the density better, but its parameters are far
worse as width estimates: it overstates the Lorentzian width by a factor
$4.4$ at $\lambda=0.25$ (still by $6\%$ at $\lambda=4$) and the Gaussian
width by between $14\%$ ($\lambda=0.25$) and a factor $4.2$ ($\lambda=4$).
The standard common-FWHM mapping is thus the better-behaved approximation
for width interpretation, with distortions of up to about $10\%$, while the exact MLE
removes the distortion entirely. The distortion is not a defect of any
particular optimizer: the pseudo-Voigt families are different parametric families, and
their width parameters are different population quantities from the
convolution widths. Fits use multistart Nelder--Mead with reported
convergence; the script prints all fitted values.

\subsection{Conditional attribution}\label{sec:redes}

\begin{figure}[htb]
\centering
\includegraphics[width=0.49\textwidth]{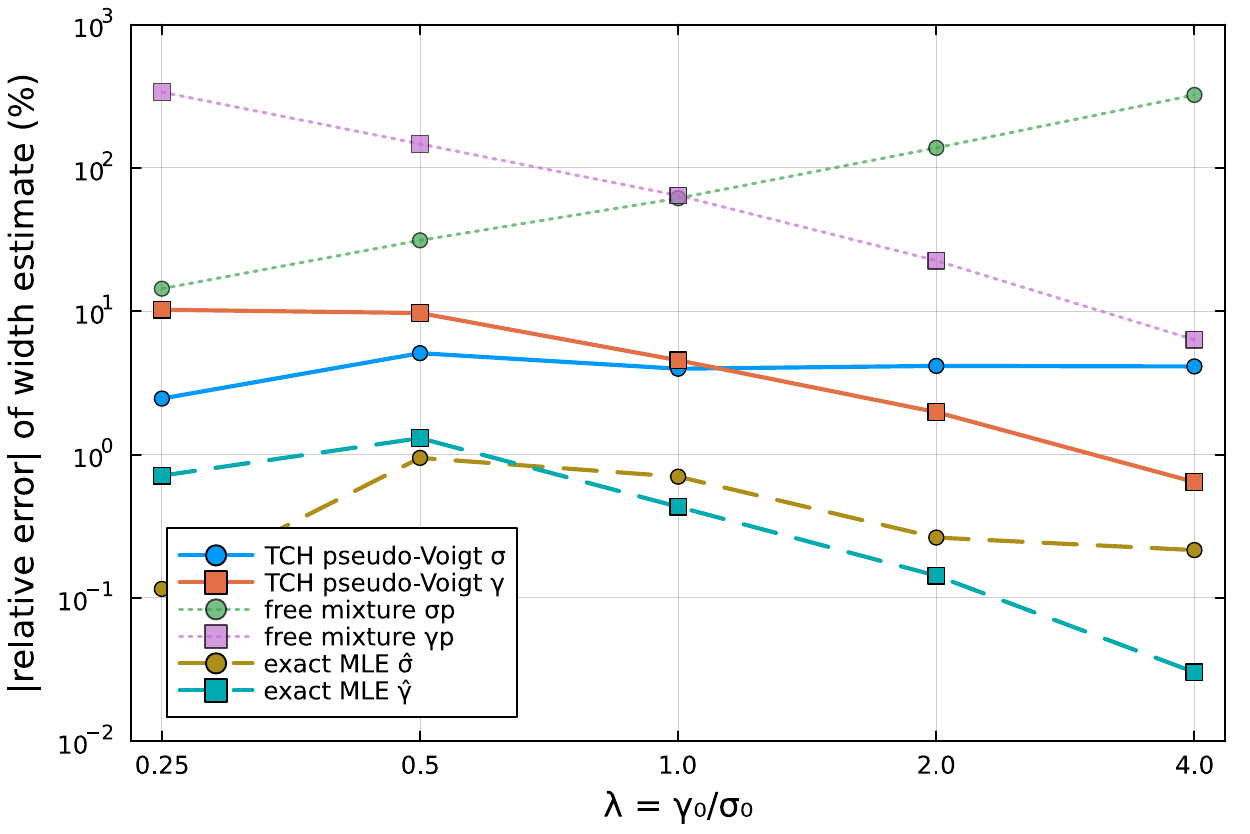}
\includegraphics[width=0.49\textwidth]{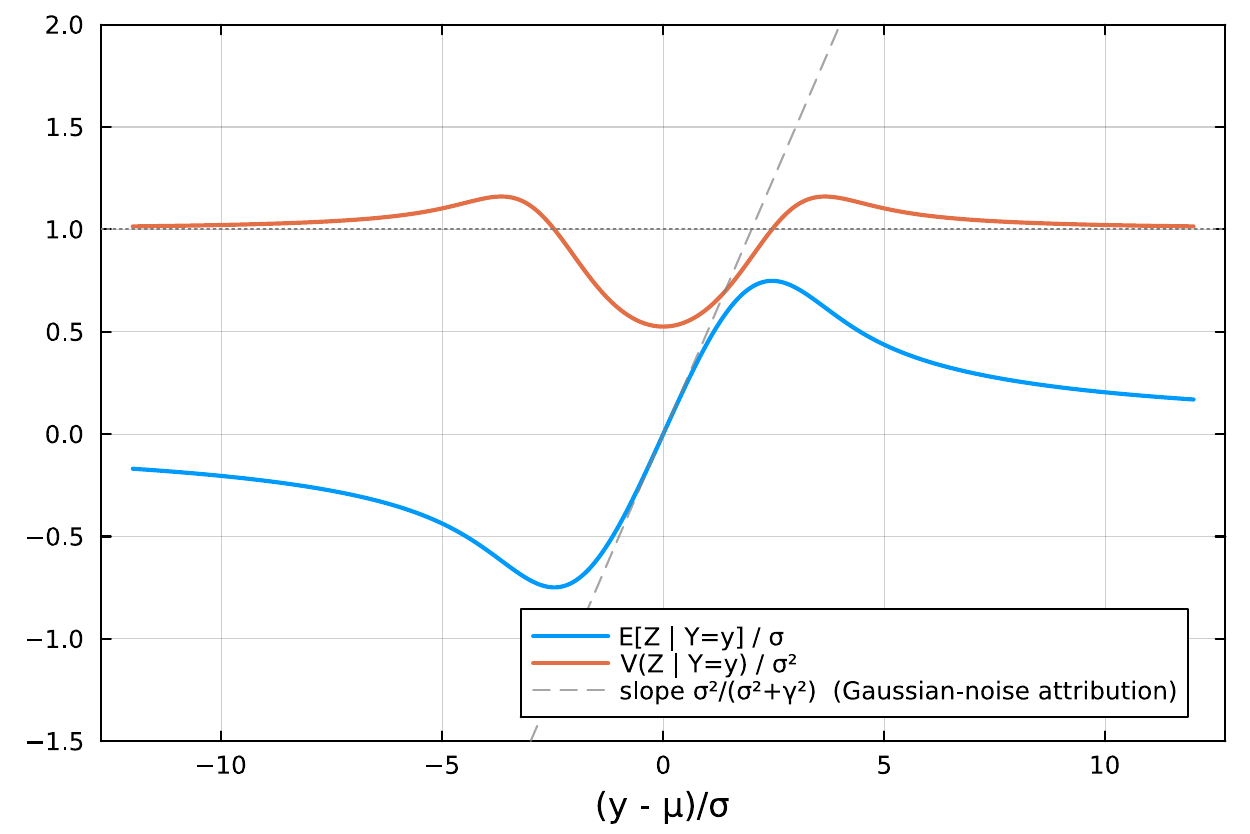}
\caption{Left: absolute relative distortion (log scale) of implied
convolution widths from the Thompson--Cox--Hastings pseudo-Voigt (solid)
and the free Gaussian--Lorentzian mixture (dotted), versus the sampling
error of the exact MLE (dashed), as a function of
$\lambda=\gamma_0/\sigma_0$; $n=2\times10^5$ per design
(Section~\ref{sec:pv}). Right: the redescending conditional moments
\eqref{eq:cond} in dimensionless units, with the Gaussian-noise benchmark
of slope $1/2$ (Section~\ref{sec:redes}). Produced by
\texttt{examples/figures.jl}.}
\label{fig:figs}
\end{figure}

The right panel of Figure~\ref{fig:figs} shows the conditional moments
\eqref{eq:cond} for $(\mu,\sigma,\gamma)=(0,1,1)$ in dimensionless units,
$\E[Z|Y=y]/\sigma$ and $\V(Z|Y=y)/\sigma^2$ against $(y-\mu)/\sigma$. The
benchmark line is the attribution that would obtain if the Lorentzian
component were replaced by \emph{Gaussian} noise $\mathcal N(0,\gamma^2)$
(the Cauchy component has no variance, so the benchmark is fixed by matching
the scale parameter): a straight line of slope
$\sigma^2/(\sigma^2+\gamma^2)=1/2$, the Kalman-type linear attribution. The
exact conditional mean has nearly the same central slope
($1-\V(Z|Y=0)/\sigma^2\approx0.475$) but redescends in the tails, where the
deviation is attributed to the Lorentzian component.

\FloatBarrier
\subsection{A measured Raman line}\label{sec:raman}

\begin{figure}[htb]
\centering
\includegraphics[width=0.8\textwidth]{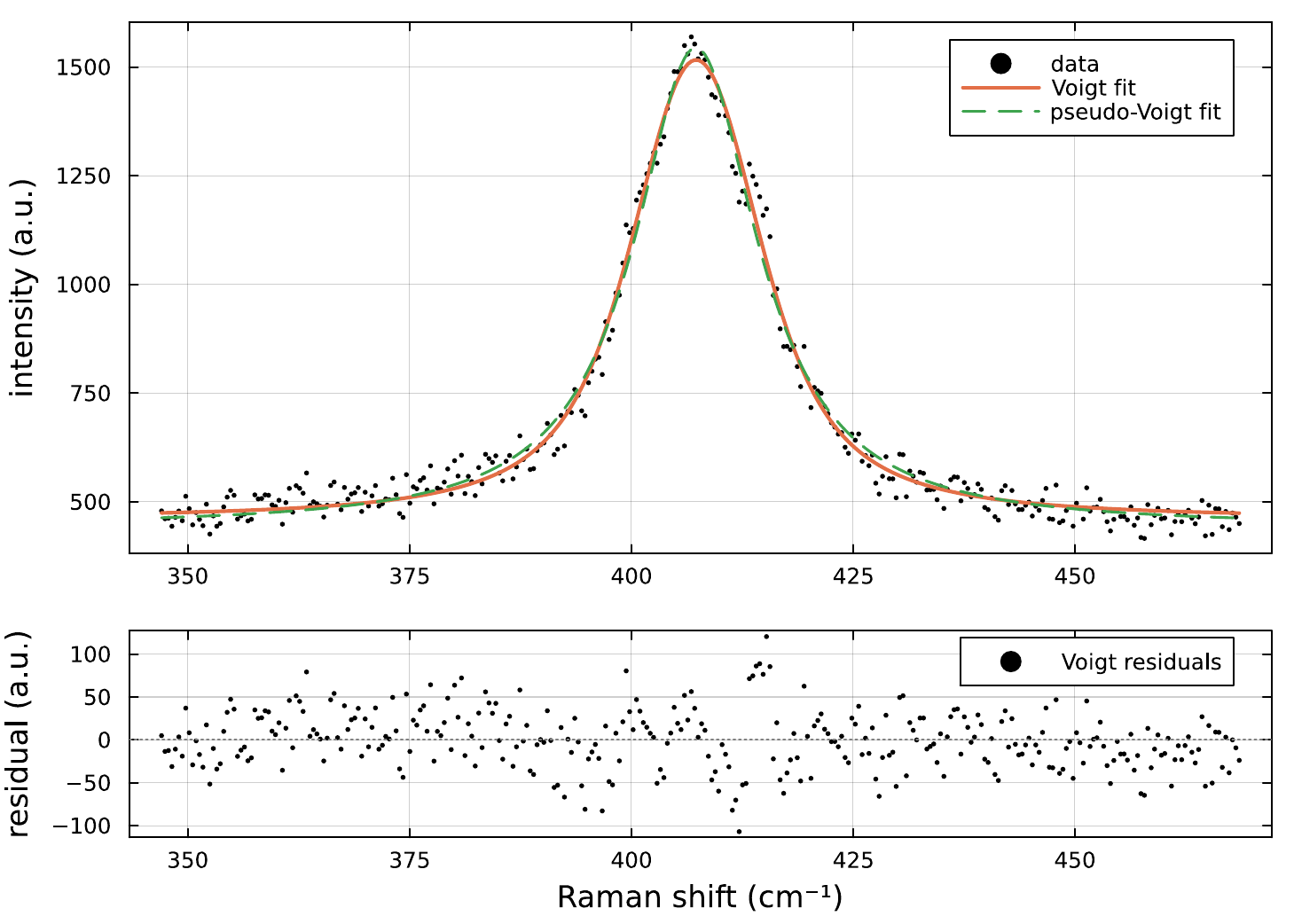}
\caption{Exact Voigt fit (analytic Jacobian) and free Gaussian--Lorentzian
mixture (pseudo-Voigt) fit of a red-ochre Raman peak ($316$ points; data
distributed with the CRAN \texttt{voigt} package \cite{CannasPiras:2025},
from Pisu et al.\ \cite{Pisu:2025}). Lower panel: residuals of the Voigt
fit. Produced by \texttt{examples/raman.jl}.}
\label{fig:raman}
\end{figure}

Figure~\ref{fig:raman} shows a fit to a measured spectrum: the red-ochre
Raman peak distributed with the CRAN \texttt{voigt} package (version 2.0)
of \cite{CannasPiras:2025}, from the provenance study of Pisu et al.\
\cite{Pisu:2025} ($316$ points; the data are read from that package's CSV
export and are not redistributed with our archive). The intensity model is
$I(\nu)=b+A\,f_Y(\nu;\mu,\sigma,\gamma)$, fitted by Levenberg--Marquardt
\cite{Marquardt:1963} with the fully analytic Jacobian
$\partial I/\partial(b,A,\theta)'=(1,\;f_Y,\;A\,f_Y\,s_\theta')'$, which
reuses the same Faddeeva evaluation as the model and makes the relevance of
the likelihood score to line refinement transparent. The fit converges in
$32$ iterations to $\hat\mu=407.310\,(0.064)$, $\hat\sigma=3.97\,(0.35)$,
$\hat\gamma=6.22\,(0.39)$, with $R^2=0.9866$; the parenthetical standard
errors are the conventional least-squares estimates
$\hat s^2(J'J)^{-1}$ evaluated at the optimum. These are local-curvature
estimates conditional on the model and on independent, homoskedastic
residuals; spectral residuals can be heteroskedastic or serially dependent,
so we present them as conventional reference values rather than as a full
uncertainty analysis. The Gaussian and Lorentzian
\emph{component} FWHMs of the fitted profile are
$w_G=\sqrt{8\ln2}\,\hat\sigma=9.35$ and $w_L=2\hat\gamma=12.44$ (we do not
attribute them to instrumental versus lifetime broadening, which a single
peak cannot separate). As a robustness check, replacing the constant
background by a linear one, $I(\nu)=b_0+b_1(\nu-\bar\nu)+A\,f_Y$, picks up
a small negative baseline slope ($\hat b_1=-0.19$) but leaves the widths
essentially unchanged ($\hat\sigma$: $3.97\to4.00$, $+0.8\%$;
$\hat\gamma$: $6.22\to6.18$, $-0.6\%$; both well within one standard
error), so the width decomposition is not compensating for a misspecified
baseline. A free Gaussian--Lorentzian mixture fit of the same
peak reaches the pure-Lorentzian boundary $\hat\eta=1$
($\hat\gamma_p=8.48$, $R^2=0.9848$), at which $\sigma_p$ is unidentified
(its Jacobian column vanishes); consequently the mixture provides no
estimate of a Gaussian convolution width, while the convolution fit
attributes a substantial Gaussian component: a real-data counterpart of the
distortion shown in Figure~\ref{fig:figs}.

\FloatBarrier
\section{Summary}

The Voigt profile admits an exact and computationally inexpensive likelihood
calculus: score, Hessian, and the conditional moments of the Gaussian component
are algebraic in $K$ and $L$ from a single Faddeeva evaluation per data point,
and the expected Fisher information requires only one-dimensional quadrature.
For fixed interior widths $\sigma,\gamma>0$, maximum likelihood is a
regular $\sqrt n$ problem despite the absence of a finite mean or
variance. The \textsf{voigtinference} package (Python, with a Julia companion that
agrees within $10^{-12}$ on the distributed validation grid) implements the full
toolkit (score, Hessian, Fisher information, maximum likelihood with
boundary diagnostics, and conditional moments), including the validated
Cauchy-limit branches that keep the score and Hessian accurate where the
plain formulas lose all significant digits. For the normalized,
constant-width Voigt model that underlies unbinned resonance fitting and
line-shape refinement workflows, pseudo-Voigt approximations, numerical
convolution, and finite-difference derivatives are unnecessary; more
general shapes (energy-dependent or relativistic widths, truncated
profiles, non-Gaussian instrument response) remain outside this calculus
and may still require numerical convolution.

\section*{Declaration of generative AI and AI-assisted technologies
in the manuscript preparation process}

The likelihood calculus for the Voigt distribution (the score, the Hessian, the
expected information, and the conditional component moments) is developed in the
companion paper by the authors, and the first implementations of maximum
likelihood estimation, in Matlab and Julia, were written by the authors without
AI assistance. The design of this note, its validation protocol, and its
conclusions are the authors' own. During the preparation of this work the authors
used large language models (Claude, Anthropic; ChatGPT, OpenAI) to assist with
the implementation of the Python and Julia packages, the derivation and tuning of
the series expansions used in the extreme-width-ratio branches, the construction
of the validation and benchmark scripts, and the drafting and editing of the
text. Every analytical result was checked against an independent
arbitrary-precision reference, and the validation script distributed with the
package checks the branch expansions on a grid spanning sixteen decades of
width ratio. Every numerical claim in the paper is reproduced by the scripts in
the public repository, and the two implementations were cross-validated against
one another. The authors reviewed and edited all output, verified it
independently, and take full responsibility for the content of the article.

\section*{Data and code availability}
The packages and the scripts reproducing all numerical tables and figures are
available at \url{https://github.com/reinhardhansen/voigtinference}. The
results reported here correspond to release v1.1.1; the immutable source
archives and their checksums are on the repository's Releases page.

\end{document}